\documentclass[twocolumn,prl]{revtex4}

\usepackage{graphicx}
\usepackage{epsfig}
\usepackage{amsmath, amssymb}
\usepackage{verbatim}

\def\ga{\gamma}
\def\Ga{\Gamma}
\def\ka{\kappa}
\def\sgn{{\rm sgn}}

\begin{document}

\title{Topological response theory of doped topological insulators}
\author{Maissam Barkeshli}
\author{Xiao-Liang Qi}
\affiliation{Department of Physics, Stanford University, Stanford, CA 94305 }

\begin{abstract}
We generalize the topological response theory of three-dimensional topological insulators (TI) to metallic 
systems -- specifically, doped TI with finite bulk carrier density and a time-reversal symmetry 
breaking field near the surface. We show that there is an inhomogeneity-induced Berry phase 
contribution to the surface Hall conductivity that is completely 
determined by the occupied states and is independent of other details such as band dispersion and 
impurities. In the limit of zero bulk carrier density, this intrinsic surface Hall conductivity reduces to the
half-integer quantized surface Hall conductivity of TI. Based on our theory we predict the behavior of 
the surface Hall conductivity for a doped topological insulator with a top gate, which can be directly 
compared with experiments.
\end{abstract}

\maketitle

Topological insulators (TI) are states of matter with a bulk energy gap and topologically protected
gapless edge states. Their most striking properties are their topological responses, such as
the quantized Hall conductance of integer quantum Hall (QH) states \cite{klitzing1980}. Recently,
a wide class of time-reversal (TR) symmetry protected topological insulators has been
discovered \cite{QZphysToday,moore2010,hasan2010,qi2010RMP}. The three-dimensional ones
are crystalline materials with a bulk energy gap and massless
linearly dispersing Dirac-like surface states \cite{moore2007,fu2007b,roy2009}. The topological
responses of 3D TI can be observed when the surface states are gapped by magnetism \cite{qi2008b}
or superconductivity \cite{fu2008}.
When the surface states are gapped by magnetism, the TI obtains a topological electromagnetic response
described by the effective action $S_\theta=\frac{\theta}{2\pi}\frac{\alpha}{2\pi} \int d^3xdt
{\bf E \cdot B}$ with $\alpha=e^2/\hbar c$ the fine structure
constant, and $\theta=\pi$ modulo $2\pi$ \cite{qi2008b}.
This is a topological magneto-electric effect and
can be observed as a surface QH effect with a half integral quantized Hall conductivity
$\sigma_{xy} = \frac{\theta}{2\pi} \frac{e^2}{h}=(n+\frac12)\frac{e^2}h$, with $n$ an integer.
For non-interacting topological insulators $\theta$ can be calculated explicitly as the
Chern-Simons invariant of the geometrical gauge field defined in momentum
space \cite{qi2008b,essin2009}.
The topological response theory of 3D TI can be obtained by a dimensional reduction
from the 4D generalization of the quantum Hall effect \cite{zhang2001}.

With the recent systematic understanding of topological insulators, a natural question
is whether only insulators can be topological: are there topological
phenomena in gapless systems such as a Fermi liquid?
In this paper, we develop a theory to characterize the intrinsic response properties in a
generic 3D Fermi liquid. 
Although in a metallic system the response properties
such as Hall conductivity are not quantized, there can still be intrinsic ``topological'' contributions
that are determined by the Berry phase gauge field in momentum space and are independent
of non-topological details such as energy dispersion away from the
Fermi surfaces and impurities.
\begin{figure}[tb]
\centerline{
\includegraphics[width=1.5in,height=1.5in]{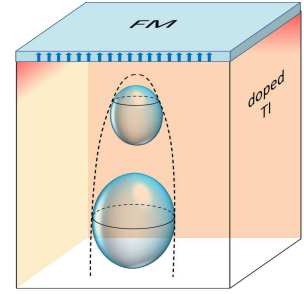}
}
\caption{Schematic picture of a doped TI with a ferromagnetic layer on the surface.
The spheres indicate the Fermi surface size which is position dependent.
Near the surface, the Fermi level lies in the gap.
\label{fig1}
}
\end{figure}

An interesting example of a 3D Fermi liquid with nontrivial topological response is a doped TI--a system with the band structure of a TI but with a finite bulk carrier density.
Here, we study the surface Hall conductivity of a doped TI with TR breaking field applied at the surface, as illustrated in Fig. \ref{fig1}.
(Some other behavior of doped TI inherited from the nearby TI has been
studied recently \cite{BR1017, HG1030}).
Unlike a TI, the surface Hall conductivity of a doped TI is not quantized, and is in
general dependent on surface conditions. However, in the smooth boundary limit, where the
changes of the Fermi level and the band structure near the boundary are smooth on the scale of the mean free
path, we show that the surface Hall conductivity is determined by a topological invariant---the Chern-Simons invariant
of the geometrical gauge field integrated over the two-dimensional Fermi surface and the real space direction
perpendicular to the boundary. We obtain our result by generalizing the dimensional reduction approach in
Ref. \cite{qi2008b} to metallic systems. Our result can be understood as a
higher dimensional, inhomogeneity-induced generalization of
the intrinsic anomalous Hall effect \cite{KL5454,JN0208,XC1059,NS1039} and its interpretation as
a topological property of Fermi liquids\cite{H0402}. Our results for doped TI are relevant to the current
experiments on TI materials since all the known 3D TI materials, such as
Bi$_{1-x}$Sb$_x$ alloy\cite{fu2007a,hsieh2008}, Bi$_2$Se$_3$ and
Bi$_2$Te$_3$\cite{hjzhang2009,xia2009,chen2009}, have finite
residual carrier density, which has become a major experimental challenge for
the observation of topological effects in TI. We also numerically studied the surface Hall conductivity
of doped Bi$_2$Se$_3$ with a simple surface T-breaking field using the four band effective model \cite{hjzhang2009,liu2010};
our results can be verified by zero-field transverse conductance measurements of Bi$_2$Se$_3$ with a
magnetic layer deposited on the surface and voltage applied at a top gate.

Consider a Hamiltonian $H_{3D}[A_\mu, \theta(\v r)]$ for a 3+1D non-interacting fermion system that is coupled to the electromagnetic gauge field $A$ and
that depends on a spatially varying parameter $\theta(\v r)$. For example, in a doped TI with a boundary, $\theta(\v r)$ may describe both the
spatial variation of the Fermi energy and the change of band structure across the boundary. Without losing generality, we can define
$\theta = 0 $ and $\pi$ to describe the vacuum and the bulk of the doped TI, respectively.
To study the surface of a doped TI, we take $\theta({\v r}) = \theta(z)$ to depend
on one direction of space, with
$\theta (z \rightarrow - \infty) = \pi$ deep in the bulk, while $\theta (0) = 0$ at the surface.
When $\theta$ varies smoothly, at each point in space the system without external magnetic field
is locally described by a Bloch Hamiltonian $H=\sum_{\bf k}c_{\bf k}^\dagger h[{\v k}, \theta(z)]c_{\bf k}$,
where ${\v k }= (k_x, k_y, k_z)$. For $N$ bands, $c_{\bf k}$ is a $N$-dimensional vector and 
$h[{\v k}, \theta(z)]$ is a $N\times N$ Hermitian matrix. In the physical system $\theta$ is defined in the region $[0,\pi]$. It is convenient to
extend the definition of $h[\v k, \theta]$ to $\theta \in [0, 2\pi]$ by setting
$h[{\v k}, \theta] = T^{\dagger}  h[-{\v k}, 2\pi - \theta] T$
for $\pi \leq \theta \leq 2\pi$. Here $T$ is the time-reversal
operator, and $h[{\v k},\theta]$ is continuous in $\theta$ since it is TR invariant at $\theta=0,\pi$.

Now we study the electromagnetic response of the system, which is determined by the effective action
\begin{eqnarray}
e^{iS_{\rm eff}[A_\mu,\theta]}=\int \mathcal{D}\bar{c}\mathcal{D}c e^{i\int dt\left[\sum_i\bar{c}_ii\partial_t c_i-H_{3D}[A,\theta]\right]} .
\end{eqnarray}
The DC Hall conductivity is determined by the quadratic term in the action:
\begin{eqnarray}
S^{\rm Hall}_{\rm eff}[A_\mu,\theta]=\frac12\int d^3xdt\epsilon^{\mu\nu\sigma\tau}\sigma_\mu[\theta] A_\nu\partial_\sigma A_\tau .
\end{eqnarray}
with the Greek letters $\mu,\nu...=0,1,2,3$ labeling the time and space coordinates. The spatial components $\sigma_i,~i=1,2,3$ of the four vector $\sigma_\mu$ are the Hall conductivity in $yz$, $zx$ and $xy$ plane, respectively;
that is, $\sigma_i = \epsilon^{ijk} \sigma_{jk}$, with $\sigma_{jk}$ the conductivity tensor. In general, $\sigma_i=\sigma_i[\theta({\bf r})]$ is a functional of the inhomogeneous $\theta({\bf r})$.
For a smooth function $\theta({\bf r})$, one can expand $\theta({\v r}) = \theta_0 + {\bf r\cdot\nabla}\theta({\v r})|_{{\bf r}=0}$ with $\theta_0=\theta({\bf r}=0)$. 
In this expansion $\sigma_i$ can be expanded to \footnote{It can be
  verified that all other terms at the same order such as
  $\sigma_i\propto \epsilon_{ijk}n_j\partial_k\theta$ (with $n_i$ a
  constant vector) are required to vanish in the DC limit by current conservation.}
\begin{eqnarray}
\sigma_i[\theta({\bf r})]\simeq C_{1i}(\theta_0)+\frac 1{2\pi}G_3(\theta_0)\partial_i\theta .
\end{eqnarray}
The coefficients in the two terms are given by two-point and three-point correlation functions, respectively. Using the time-ordered Green's function $G({\v k},\omega)=\left[\omega-e^{-i\delta}h({\v k})\right]^{-1}$, the coefficients can be expressed as:
\begin{eqnarray}
C_{1i}(\theta_0)&=& \frac{1}{8}\int\frac{d^3k d\omega}{(2\pi)^3} \epsilon^{ijkl} {\rm Tr} \left[ G \partial_j G^{-1} G \partial_k G^{-1} G \partial_l G^{-1}\right]\nonumber\\\label{C1}
\\
G_3(\theta_0) &= &-\frac{i\pi^2}{3} \epsilon^{ijkl} \int \frac{ d^3k d\omega}{(2\pi)^5}
{\rm Tr}\left[G\partial_iG^{-1}G\partial_jG^{-1}\right.
\nonumber \\
&&\cdot\left. G\partial_kG^{-1}G\partial_lG^{-1}G\partial_\theta G^{-1} \right].\label{G3}
\end{eqnarray}
with $\partial_j=\partial/\partial k_j$ and $k_j$ is the frequency and momentum vector with $k_0=\omega$ the frequency and $(k_1,k_2,k_3)={\v k}$ the momentum.

Before evaluating the correlation functions, let us discuss the criteria for the
validity of the above effective theory approach.
In the case of a band insulator with a gap $E_g$, there is a length scale $\xi=\hbar v/E_g$ where
$v$ is a typical velocity scale contained in the
coupling between conduction and valence bands. The approach above is valid as long as the characteristic scale of variations of $\theta$
is much longer than $\xi$, {\it i.e.}, $|\nabla \theta|\ll 1/\xi$. For a perfect metal, the gap vanishes and the Green's function in
(\ref{C1}) and (\ref{G3}) is singular. With impurities, the Green's function obtains a finite self-energy $1/\tau$, which corresponds
to a length scale $l=v_F\tau$, which is the mean-free path. The mean-free path plays the role of $\xi$ in the insulator case, and
the expansion above applies to the system with $|\nabla \theta|\ll 1/l$. Physically, the expansion applies because the conductivity at position $\v r$ will be insensitive to small changes in $\theta$ that occur at $\v r'$ when $|\v r' - \v r| >> l$. In other words, the intrinsic contribution to the Hall
conductivity calculated above applies in the ``dirty limit" where the Hamiltonian is smoothly varying on the
scale of the mean-free path, 
although the value of the intrinsic Hall conductivity is independent of the impurity strength in this limit.
In general, the anomalous Hall conductivity of a metal also receives extrinsic contributions arising from impurity scattering,
which are termed the side-jump and skew-scattering contributions \cite{NS1039}. Here we focus on
the intrinsic contributions, which depend only on the nature of the Bloch states and their associated Berry phases. In the
case of ferromagnetic metals, the intrinsic contribution can dominate the full anomalous Hall conductivity in certain
intermediate regimes of disorder \cite{NS1039}; thus we expect that the intrinsic contribution that we calculate here for
the surface of a doped TI may also be the dominant contribution in certain regimes of disorder.

Under the condition discussed above, the gradient expansion can be done locally around each point in space, leading to a local Hall conductivity.
For simplicity, we consider $\theta({\bf r})=\theta(z)$ only varying along the $z$ direction. The Hall conductivity in the $xy$ plane is given by
$\sigma_{z}(z) = C_1^z[ \theta(z)] + \frac{G_3}{2\pi}[\theta (z)] \partial_z \theta(z)$.
As shown in Fig. \ref{fig1}, we consider a single interface at $z = 0$ between the vacuum at $z>0$ and a doped topological
insulator at $z<0$. One can always take $\theta=-\pi$ for the bulk doped TI and $\theta=0$ for the vacuum, both of which are TR invariant. In such a situation, the total current in the $xy$-plane will be given by
$j^\mu_{2D} = \int dz j^\mu $, from which we may obtain the two-dimensional Hall conductivity
$\sigma_{z}^{2D} = \int dz \sigma_{z}(z) = \sigma_{z}^{2D;(1)}  + \sigma_{z}^{2D;(2)} $. The first term
$\sigma_{z}^{2D;(1)} =\int dz C_{1z}[\theta_0(z)] = \frac{1}{4\pi^2} \sum_{\alpha} \int d^3k dz \mathcal{F}_{xy}^{\alpha}(z) n_\alpha (\v k,z)$
is the intrinsic anomalous Hall conductivity of a homogeneous system\cite{KL5454,JN0208}, where $\alpha$ labels the occupied bands $\left|\alpha{\bf k}\right\rangle$, and $n_\alpha$ and $\mathcal{F}^\alpha_{xy}$ are the occupation number and Berry curvature of band $\alpha$. More explicitly $\mathcal{F}^\alpha_{xy}=\partial_x A_y^\alpha-\partial_yA_x^\alpha$ with $A_i^\alpha=-i\left\langle \alpha {\bf k}\right|\partial_{k_i}\left|\alpha{\bf k}\right\rangle$. 
Modulo a quantized contribution $n/2\pi$ with integer $n$, $\sigma_{z}^{2D;(1)}$ is determined by the Berry phase gauge field at the Fermi surface\cite{H0402}. 
The second term
$\sigma_{z}^{2D;(2)} = \frac{1}{2\pi} \int dz G_3[\theta_0(z)] \partial_z \theta = \frac{1}{2\pi}\int_{-\pi}^{0}G_3(\theta) d\theta$
is the new contribution induced by spatial inhomogeneity.
With TR symmetry only broken at the interface, one can extend the definition of $\theta$ to $[0,2\pi]$
as mentioned above. By TR symmetry it can be proved that $G_3(\theta)=G_3(-\theta)$, so that $\int_0^\pi d\theta G_3(\theta) = \frac{1}{2} \int_0^{2\pi} d\theta G_3(\theta)$.

Therefore, as long as the start and end points are described by time-reversal invariant Hamiltonians,
we find that the integrated current response receives a contribution from:
\begin{align}
\label{j2D}
\sigma_{xy}^{2D;(2)} \equiv \sigma_{z}^{2D;(2)} = \frac{e^2}{h} \frac12 f,
\end{align}
where $f = \int_0^{2\pi} G_3(\theta) d \theta$, and we have recovered the units $e^2/h$. From the definition of $G_3$ in Eq. (\ref{G3}) one can see that $\theta$ plays exactly the same role as momentum and frequency. The integrand in Eq. (\ref{G3}) has a symmetric form if we view $\theta$ as a momentum component in an ``extra dimension."
In the case of an insulator, $f$ is a topological invariant which is equal to the second Chern number associated
with the Bloch states of the 4+1D Hamiltonian $H[\v k, \theta]$. For a
TI, $f$ will be odd, so one obtains a half-integral
quantized transverse conductivity at the surface of a TI,
when the surface is fully gapped \cite{qi2008b}.

\begin{figure}[tb]
\centerline{
\includegraphics[scale=0.16]{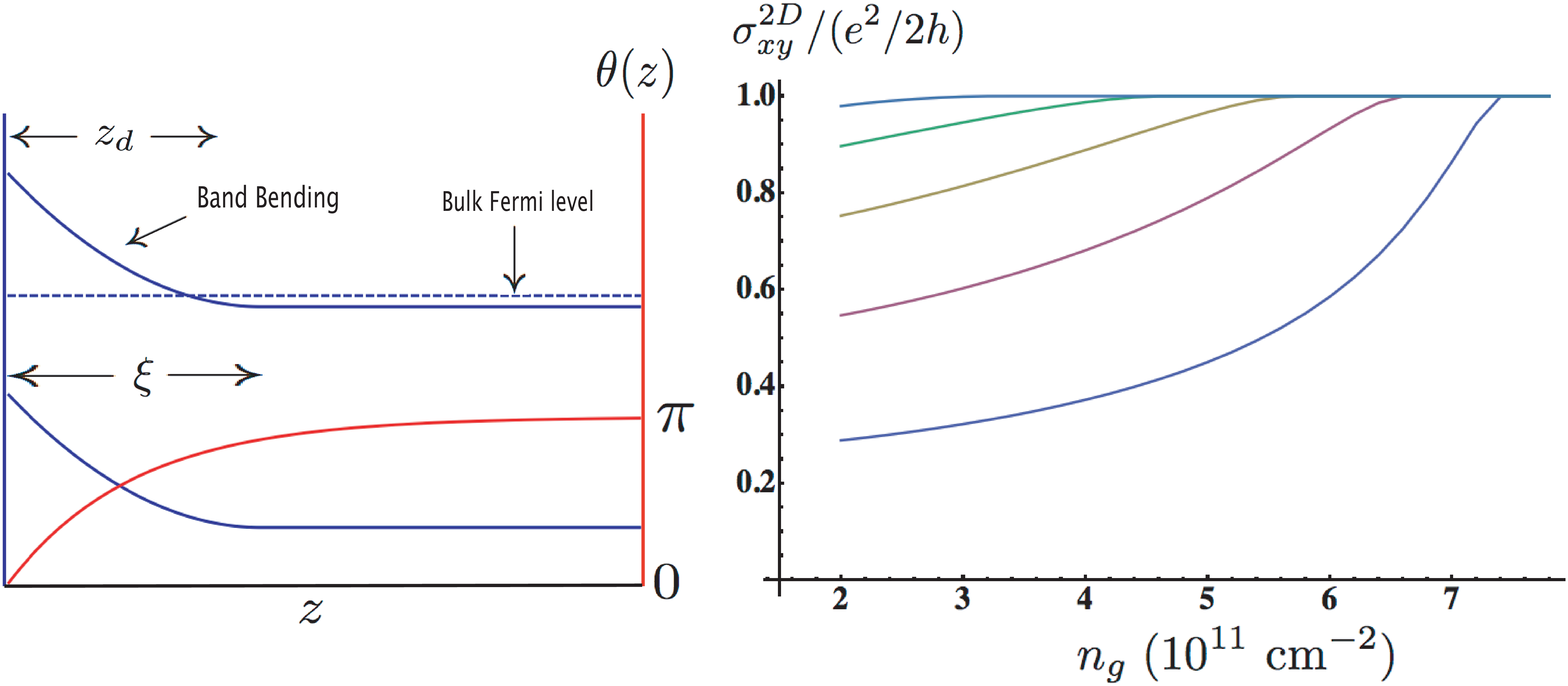}
}
\caption{
\label{images}
Left: blue indicates schematically the band bending, displaying metallic bulk but gapped surface.
Red indicates variation of $\theta$ from 0 to $\pi$.
Right: numerical results for $\sigma_{xy}^{2D}$ as a function of the top gate carrier
density $n_g = C V_g/e$. The penetration depth of the T-breaking perturbation is chosen
to be $\xi = 1$ nm, the depletion length is $z_d = n_g/n_b$,
the bulk carrier density $n_b = 10^{16} cm^{-3}$, and the Fermi level
is chosen to be 18 meV from the conduction band bottom \cite{AC1007}. Starting from the bottom, the different
curves correspond to $m_0= 50$, $100$, $150$, $200$, $250$ meV, respectively. The gate carrier densities
$n_g \sim 10^{11} cm^{-2}$ are within experimentally accessible values. 
}
\end{figure}

For a doped TI, the Green's functions have poles on the Fermi surface, so that $f$ is in general not quantized.
However, $f$ is still invariant upon arbitrary smooth deformations of the Green's function $G({\v k},\theta,\omega)$ as
long as the deformation vanishes at the Fermi surface. Making use of such topological invariance, we are able to integrate over
frequency $\omega$ and obtain the following expression for $f$ (see Supplementary Materials for the proof):
\begin{eqnarray}
 f&=&  \frac{1}{32\pi^2}\int_{BZ} d^3{\bf k}\int_0^{2\pi}d\theta{\rm  Tr}\left( \mathcal{F}_{ab} \mathcal{F}_{cd}\right) \epsilon^{abcd}\label{C2result} .
\end{eqnarray}
Here $\mathcal{F}_{ab}$ is the Berry curvature associated with the bands that are filled, with $a,b=1,2,3,4$ labeling $k_x,k_y,k_z,\theta$. More explicitly 
$\mathcal{F}_{ab}^{\alpha \beta} = \partial_a \mathcal{A}_b^{\alpha \beta} -  \partial_b \mathcal{A}_a^{\alpha \beta}
+ i [\mathcal{A}_a, \mathcal{A}_b]^{\alpha \beta}$, $i \mathcal{A}_a^{\alpha\beta} =  \langle \alpha {\v k} \theta | \partial_a | \beta {\v k} \theta \rangle$ with $\alpha,\beta$ running  over filled bands.
Formally, (\ref{C2result}) looks the same as the second Chern number of the Berry phase gauge field that appears in
the TI case. However, for metallic systems there are different numbers of filled bands inside and outside each Fermi surface,
so that the definition of $\mathcal{F}_{ab}$ depends on the position
of the Fermi surface and therefore $f$ is {\it not} a second Chern number. For example, if there is a single
Fermi surface, then $32\pi^2  f =  \int_{FFS} Tr \mathcal{F}^{(1)} \wedge \mathcal{F}^{(1)} + \int_{BZ \backslash FFS} Tr \mathcal{F}^{(0)} \wedge \mathcal{F}^{(0)}$,
where $FFS$ denotes the filled Fermi sea, $\mathcal{F}^{(1)}$ is the Berry curvature associated with
filled bands inside the Fermi surface, and $\mathcal{F}^{(0)}$ is the curvature associated with
filled bands outside of the Fermi surface.

Since the second Chern form is a total derivative, $f$ can be expressed, modulo an integer, in terms of integrals over
the Fermi surfaces. We find that the fractional part of $f$ is:
\begin{align}
\label{C2frac}
f \text{ mod } 1 = \frac{1}{8 \pi^2}\sum_i \int_{FS_i} \mathcal{L}_{CS}(\mathcal{A}^{P_i}),
\end{align}
where $\mathcal{L}_{CS}(\mathcal{A}) = \epsilon^{ijk} Tr [ \mathcal{A}_i \partial_j \mathcal{A}_k + \frac{2i}{3} \mathcal{A}_i \mathcal{A}_j \mathcal{A}_k]$ is the Chern-Simons form.
The sum is over the different Fermi surfaces, where $FS_i$ labels the $i$th three-dimensional
phase-space Fermi surface, and $\mathcal{A}^{P_i}$ is the Berry connection associated with the
partially filled bands that are crossing the $i$th
Fermi surface, \it i.e. \rm that have zero energy for $(\v k,\theta) \in FS_i$. Eq. \ref{C2frac}
is a consequence of a highly non-trivial cancellation of cross-terms associated with the
non-Abelian CS forms, which is why only the partially filled bands participate. The proof of
Eq. (\ref{C2result}) and (\ref{C2frac}) is given in the Supplementary Materials.

Eq. (\ref{C2frac}) is an integral over the Berry connection involving only states at the Fermi surface. This is in
direct analogy to Haldane's result \cite{H0402} that the fractional part of the intrinsic contribution to the anomalous Hall effect
is due to the adiabatic phase acquired by quasiparticles propagating on the Fermi surface. We see that
inhomogeneities can also induce Berry phase contributions to the anomalous Hall effect, the fractional part of
which depends only on states at the Fermi surface.

Note that in order for the above results to hold, we have assumed that $\theta$ varies slowly on the scale set by the
mean free path. We must also assume that the Fermi surface vanishes near the surface of the
material (see Figs. \ref{fig1} and \ref{images}); otherwise, the surface will cause a sharp jump in the Fermi level and cannot
be treated as a smooth change. If the Fermi level already lies in a gap at the surface, then the sharp jump in the chemical
potential will not have consequences for the low energy physics.

Let us now apply our result to calculate the surface Hall conductivity of a realistic doped TI---doped Bi$_2$Se$_3$. 
From the discrepancy between bulk transport measurements like Shubnikov-de Haas oscillations and surface 
measurements like ARPES, it has been noticed that the energy bands of Bi$_2$Se$_3$ (and Bi$_2$Te$_3$) can 
bend upwards near the surface, so that the Fermi level lies in the energy gap near the surface but in the conduction 
band in the bulk of the material (see Fig. \ref{images}) \cite{AC1007}. Such a situation can be further tuned
using a top gate voltage. This can be described by our current approach since the gap near the surface effectively avoids the abrupt change in the band 
structure at the surface. The bulk of the system is TR invariant and can be described by the four-band 
effective model \cite{hjzhang2009,LQ1022} $H_0[{\v k}] = \epsilon({\v k}) \mathbb{I} + \sum_{a} d_a^{(0)}({\v k}) \Ga^a$ 
with $\Gamma^a,~a=0,1,2,3$ the Dirac matrices and all the parameters $d_a^{(0)}({\v k})$ and $\epsilon({\v k})$ given in 
Ref. \cite{LQ1022} (see also the Supplementary Materials).

We consider a T-breaking term induced by, {\it e.g.}, magnetic dopants deposited at the surface.
Recently, angle-resolved photon emission (ARPES) experiments have
observed a gap in the surface states of ${\rm Mn}$ doped ${\rm
  Bi_2Te_3}$ \cite{chen2010c}, with the surface Fermi level lying in the gap. This system may provide an
experimental realization of the surface TR symmetry breaking, although further verification of
the nature of the surface gap is still needed. This can be described by adding the following surface term to the Hamiltonian:
\begin{align}
H_1 = eV(z) + m_0 \sin [\theta(z)] \Ga^3 - M_0 (1 + \cos[\theta(z)]) \Ga^5,
\end{align}
where $eV(z) = - \frac{eC V_g}{4 \epsilon \epsilon_0} z_d (1 - \frac{z}{z_d})^2$ is the potential induced by
the top gate voltage, $C$ is the capacitance per unit
area of the dielectric separating the top gate and the surface of the TI, $n_b$ is the bulk carrier density of the 
TI, $V_g$ is the voltage applied at the top gate, $z_d = CV_g/en_b$ is the depeletion width induced by the top gate,
and $\epsilon$ is the relative dielectric constant of the TI \cite{CXLiu} (see Supplementary Materials). Note that charge neutrality 
requires $z_d$ to be determined by the gate voltage, which reduced the 
number of free parameters in our calculation, allowing us to determine the Hall conductivity from realistic material parameters.

We assume that the band bending near the surface is entirely induced by the top gate. 
$\theta(z) = \pi (1 - e^{-|z|/\xi})$ describes the surface TR symmetry breaking which decays exponentially versus 
the perpendicular distance $z$. The penetration depth $\xi$ is set by the thickness of the layer of Mn dopants deposited at
the surface; we take $\xi \approx 1 nm$, assuming they penetrate several atomic layers into the surface. 
In a real system the T-breaking term may vary depending on details of the system. 
Here we take a simple T-breaking term.
The Hamiltonian $H=H_0+H_1$ can still be written in the Dirac form 
$H[\theta(z), \v k] = (\epsilon(\v k) - \mu + eV(z))\mathbb{I} + \sum_a d_a(\v k, \theta) \Ga^a$. 
All the $z$ dependence can be absorbed into the parameter $\theta$ by defining $z(\theta) = -\xi \ln|1-\theta/\pi|$.
$d_a({\bf k}, \theta) = d^{(0)}_a({\bf k}, \theta) + \delta_{a,3} m_0 \sin [\theta] - \delta_{a,5} M_0 (1 + \cos[\theta(z)])$.
For this model Hamiltonian, $\sigma_{xy}^{2D;(1)}$ vanishes. $f$ in Eq. (\ref{C2result}) can be shown (see Supplementary Materials) to be given by
\begin{align}
f = -\frac{3}{8 \pi^2}\int_{\overline{FFS}} d^3 kd\theta
\epsilon_{tabcd} \hat{d}_t\partial_x \hat{d}_a \partial_y \hat{d}_b \partial_z \hat{d}_c \partial_\theta \hat{d}_d,
\end{align}
where the integral is carried over the outside of electron Fermi pocket, denoted by $\overline{FFS}$ and $\hat d = d/|d|$.
As discussed earlier, $\theta$ can be viewed as the fourth momentum, playing the same role as the
three-dimensional momenta $k_{x,y,z}$, and the three-dimensional phase space Fermi surface can be defined in the ${\v k},\theta$ space.

Numerical results of the surface Hall conductivity are shown in Fig. \ref{images} as a function of the 
surface gate carrier density, $n_g = C V_g$, for different strengths of the gap induced by the magnetization, $m_0$. 
Note our theory is only valid when the surface is gapped. 
The Hall conductivity saturates to the quantized value $e^2/2h$ in the limit that $z_d \approx n_g/n_b \gg \xi$. Physically 
this is because the T-breaking, and thus the nonzero Hall conductivity, only occurs in the range of 
$z\lesssim \xi$. For $\xi\ll z_d$ the system is insulating in the range of $z\lesssim \xi$, so that the 
surface Hall conductivity remains at the quantized value of TI surface. As expected, in the limit of low 
carrier density 
the Hall conductivity also approaches $e^2/2h$, which is dramatically different from an ordinary 
semiconductor in which the intrinsic Hall conductivity always vanishes in the limit of zero carrier 
density. 

If observed, such a gate tunability of Hall conductivity in a {\it bulk} crystal provides a unique signature of 
doped TI compared to ordinary bulk 
semiconductors. We note that while disorder may add extrinsic contributions to the surface Hall conductance
that we calculate here, these contributions vanish as the gate voltage is increased and the quantized
$e^2/2h$ value is recovered. Our work provides quantitative estimates for achieving the quantized value,
and by comparing with measurements, can be used as a way to estimate the strength of extrinsic contributions
to the surface Hall conductance. Following the approach of Ref. \cite{qi2008b} for topological insulators, 
our results can also be extended to general dimensions. Such generalizations lead to a framework 
to study topological phenomena in Fermi liquids in generic dimensions, which we investigate in future work. 

{\it Acknowledgement.---} We acknowledge S.C. Zhang and C.X. Liu for insightful discussions. 
This work is supported by Alfred P. Sloan Foundation (X.L.Q.) and the Simons Foundation (M.B.).
After finishing this work, we became aware that D. Bergman has made related observations.\cite{bergman2011}

\def\eqa{7}
\def\eqb{8}
\def\eqc{5}

\def\ga{\gamma}
\def\Ga{\Gamma}
\def\ka{\kappa}
\def\sgn{{\rm sgn}}

\begin{widetext}

\section{Supplementary Online Material}


The purpose of these Supplementary Materials is to prove the following results (eqns. (\eqa) and (\eqb) of the main text):
\begin{align}
\label{F1}
f &\equiv -\frac{i \pi^2}{15} \epsilon^{\mu \nu \rho \sigma \tau} \int \frac{ d^4k d\omega}{(2\pi)^5}
Tr[ G \frac{\partial G^{-1}}{\partial q^\mu} G \frac{\partial G^{-1}}{\partial q^\nu} G \frac{\partial G^{-1}}{\partial q^\rho}
G \frac{\partial G^{-1}}{\partial q^\sigma} G \frac{\partial G^{-1}}{\partial q^\tau}  ]
\nonumber \\
& =  \frac{1}{32\pi^2}\int_{BZ} d^3{\bf k}\int_0^{2\pi}d\theta{\rm  Tr}\left( \mathcal{F}_{ab} \mathcal{F}_{cd}\right) \epsilon^{abcd},
\end{align}
and
\begin{align}
\label{F2}
f \text{ mod } 1 = \frac{1}{8 \pi^2}\sum_i \int_{FS_i} \mathcal{L}_{CS}(\mathcal{A}^{P_i}).
\end{align}
The quantities $\mathcal{A}$ and $\mathcal{F}$ are defined in the text and will also be defined below.
$q = (\omega, {\bf k})$ is a five-vector including the
four-component spatial momentum ${\bf k}$ and frequency $\omega$,
and $G = \sum_{\alpha} \frac{P_\alpha}{\omega - \xi_\alpha + i \delta \sgn(\xi_{\alpha})}$ is the non-interacting single-particle
Green's function; $P_\alpha$ is a projector onto band $\alpha$ and $\xi_\alpha = \epsilon_\alpha - \mu$, where $\mu$ is the
chemical potential. For a metal, $f$ is not quantized; despite appearances, it is not a second Chern number because the
definition of the matrix $\mathcal{F}$ is such that its dimension changes at different points in momentum space,
depending on the location of the Fermi surface(s).

The sections below are organized as follows. We begin by first manipulating the expression in terms of Green's functions into a more useful form.
Next, we point out that even though $f$ is not invariant under arbitrary deformations of the band structure,
it is invariant under a certain class of deformations that we discuss. Following this, we prove some mathematical
identities that the Berry connections of the Bloch bands must generally satisfy. Finally, after developing these ideas,
we show how to use them to prove eqns. (\ref{F1}) and (\ref{F2}). We end with a discussion of the calculation of $f$ in the simpler
case of the Dirac models (Eq. (9) of the main text), which have
additional symmetries. 

\section{Preliminary Manipulations of $f$}

We have
\begin{align}
G^{-1}({\bf k},\omega) &= \omega - H_{{\bf k}} + i \delta \sgn(H_{{\bf k}}),
\nonumber \\
\partial_\omega G^{-1} &= 1
\nonumber \\
\partial_i G^{-1} &= \partial_i H_{{\bf k}} \rightarrow \langle \alpha | \partial_i H_{{\bf k}} | \beta \rangle =
\partial_i \xi_\alpha({\bf k}) \delta_{\alpha \beta} + (\xi_\beta - \xi_\alpha) \langle \alpha| \partial_i |\beta \rangle
\nonumber \\
G({\bf k}, \omega) &= \sum_\alpha \frac{ P_\alpha}{\omega - \xi_\alpha + i \delta  \sgn(\xi_\alpha) },
\end{align}
where $\partial_i = \frac{\partial}{\partial k_i}$ is a derivative with respect to a momentum coordinate.
Inserting a complete set of states for each ${\bf k}$, we have:
\begin{align}
f = -\frac{i \pi^2}{3} \frac{1}{(2\pi)^5} \epsilon^{ijkl} \int d^4k d\omega \sum_{\alpha \beta \gamma \delta}
A^{\alpha \beta}_i A_j^{\beta \gamma} A_k^{\gamma \delta} A_l^{\delta \alpha}  d_\alpha^2 d_\beta d_\gamma d_\delta,
\end{align}
where we have set $A^{\alpha \beta}_i = \langle \alpha | \partial_i G^{-1} | \beta \rangle$ and
$d_\alpha^{-1} = \omega - \xi_\alpha + i \delta \sgn(\xi_\alpha)$.

Now observe that the above sum vanishes whenever at least any three of $\alpha$, $\beta$, $\gamma$, $\delta$
are equal. For example if $\beta = \gamma = \delta$, then the summand is
$d_\alpha^2 d_\beta^3 A^{\alpha \beta}_i A^{\beta \beta}_j A_k^{\beta\beta} A_l^{\beta \alpha}$,
which vanishes upon contraction with the epsilon tensor. Similarly, if $\alpha = \beta = \gamma$,
the summand is $d_\alpha^4 d_\delta A^{\alpha \alpha}_i A^{\alpha \alpha}_j A^{\alpha \delta}_k A^{\delta \alpha}_l$, which also vanishes
upon contraction with the epsilon tensor.

Thus we are left with 3 types of terms:
\begin{align}
\label{C2sum}
f = -\frac{i \pi^2}{3} \frac{1}{(2\pi)^5} \epsilon^{ijkl} \int d^4k d\omega \left( \sum_{\alpha \beta \gamma \delta}' +
\sum_{\alpha \beta \gamma \delta}'' + \sum_{\alpha \beta \gamma \delta}''' \right)
A^{\alpha \beta}_i A_j^{\beta \gamma} A_k^{\gamma \delta} A_l^{\delta \alpha}  d_\alpha^{2} d_\beta d_\gamma d_\delta
\end{align}
In $\sum_{\alpha \beta \gamma \delta}'$, the sum is over states $(\alpha, \beta, \gamma, \delta)$ that are all different.
In $\sum_{\alpha \beta \gamma \delta}''$, the sum is over states where exactly two of the indices are equal and the others
are all different. Finally, $\sum_{\alpha \beta \gamma \delta}'''$ is over states where two pairs of indices take the same value
(e.g. $\alpha = \beta$ and $\gamma = \delta$ but $\alpha \neq \gamma$).

Let us first show that the last sum gives zero contribution. Its contribution is proportional to:
\begin{align}
\int d^4k d \omega \sum_{\alpha \beta \gamma \delta}
(\delta_{\alpha \beta} \delta_{\gamma \delta} + \delta_{\alpha \gamma} \delta_{\beta \delta} + \delta_{\alpha \delta} \delta_{\beta \gamma})
\epsilon^{ijkl} A^{\alpha \beta}_i A_j^{\beta \gamma} A_k^{\gamma \delta} A_l^{\delta \alpha} d_\alpha^2 d_\beta d_\gamma d_\delta
\end{align}
It is easy to see that the second term vanishes because of the epsilon tensor, and
that the first and third terms cancel each other after indices are relabelled (to see this,
note that $\epsilon^{lijk} = - \epsilon^{ijkl}$, since cyclic permutations have odd signature
in even dimensions).

Therefore, we see that the third sum in (\ref{C2sum}) vanishes. Now consider the second sum in (\ref{C2sum}), which we will call $B$:
\begin{align}
B \equiv -\frac{i \pi^2}{3} \frac{1}{(2\pi)^5} \sum^{''}_{\alpha \beta \gamma \delta}  \epsilon^{ijkl} \int d^4k d\omega
A^{\alpha \beta}_i A_j^{\beta \gamma} A_k^{\gamma \delta} A_l^{\delta \alpha} d_\alpha^2 d_\beta d_\gamma d_\delta
\end{align}
Recall that $\sum _{\alpha \beta \gamma \delta}^{''}$ is a sum over states where exactly two of the indices are equal
and the rest are all different. There are exactly 6 ways this can happen; let us write each one out:
\begin{align}
B = B_{\alpha  = \beta} + B_{\alpha = \gamma} + B_{\alpha = \delta} + B_{\beta = \gamma} + B_{\beta = \delta} + B_{\gamma = \delta},
\end{align}
where the subscripts indicate which terms in the sum are to be set equal to each other. We find that
$B_{\alpha = \beta} + B_{\alpha = \delta} = 0$. We also find
\begin{align}
B_{\alpha = \gamma} + B_{\beta = \delta} &=
-\frac{i \pi^2}{3} \frac{1}{(2\pi)^5} \epsilon^{ijkl}\sum_{\alpha \beta \delta}^{'} \int d^4k d\omega
A_i^{\alpha \beta} A_j^{\beta \alpha} A_k^{\alpha \delta} A_l^{\delta \alpha} d_\alpha^2 d_\beta d_\delta (d_\alpha - d_\beta).
\nonumber \\
B_{\beta = \gamma} + B_{\gamma = \delta} &=
-\frac{i \pi^2}{3} \frac{1}{(2\pi)^5} \epsilon^{ijkl} \sum_{\alpha \beta \gamma}^{'}
\int d^4k d\omega (A_i^{\alpha \beta} A_j^{\beta \beta} A_k^{\beta \gamma} A_l^{\gamma \alpha} - c.c.)
d_\alpha^2 d_\beta^2 d_\gamma
\end{align}
$\sum^{'}_{\alpha \beta \gamma}$ indicates a sum over states where $\alpha$, $\beta$, $\gamma$ are all different.
$c.c.$ denotes complex conjugation. Note that $(A^{\alpha \beta}_i)^* = -A^{\beta \alpha}_i$ if $\alpha \neq \beta$
while $A^{\alpha \alpha}_i = \partial_i \xi_\alpha$ is real.

Therefore, we see that $f$ is given by the following sum of three terms:
\begin{align}
\label{C2simple}
f &= -\frac{i \pi^2}{3} \frac{1}{(2\pi)^5} \int d^4k (x + y + z)
\nonumber \\
x({\bf k}) &=  \epsilon^{ijkl} \sum_{\alpha \beta \gamma \delta}^{'}
\xi_{\alpha \beta} \xi_{\beta \gamma} \xi_{\gamma \delta} \xi_{\delta \alpha}
\mathcal{A}^{\alpha \beta}_i \mathcal{A}_j^{\beta \gamma} \mathcal{A}_k^{\gamma \delta} \mathcal{A}_l^{\delta \alpha}
\int d\omega  d_\alpha^{2} d_\beta d_\gamma d_\delta
\nonumber \\
y({\bf k}) &= \epsilon^{ijkl} \sum_{\alpha \beta \delta}^{'}
\xi_{\alpha \beta}^2 \xi_{\alpha \delta}^2
\mathcal{A}_i^{\alpha \beta} \mathcal{A}_j^{\beta \alpha} \mathcal{A}_k^{\alpha \delta} \mathcal{A}_l^{\delta \alpha}
\int d\omega   d_\alpha^2 d_\beta d_\delta (d_\alpha - d_\beta).
\nonumber \\
z({\bf k}) &= -i \epsilon^{ijkl} \sum_{\alpha \beta \gamma}^{'}
\partial_i \xi_\beta \xi_{\beta \alpha} \xi_{\gamma \beta} \xi_{\alpha \gamma} (\mathcal{A}_j^{\alpha \beta} \mathcal{A}_k^{\beta \ga} \mathcal{A}_l^{\ga \alpha} +
\mathcal{A}_j^{\alpha \ga} \mathcal{A}_k^{\ga \beta} \mathcal{A}_l^{\beta \alpha})
\int d\omega  d_\alpha^2 d_\beta^2 d_\gamma ,
\end{align}
where we have set $i \mathcal{A}_i^{\alpha \beta} = \langle \alpha; {\bf k}| \partial_i |\beta; {\bf k} \rangle$
and $\xi_{\alpha \beta} \equiv \xi_\alpha - \xi_\beta$.
The important point to note here is that
$x({\bf k})$ and $y({\bf k})$ do not depend on the slope of the the dispersion, $\partial_i \xi_\alpha$,
while $z({\bf k})$ does.

Note that in the sum in $z({\bf k})$, all three bands $\alpha$, $\beta$ and $\gamma$
must have different energies. 
%
Consequently, for a special class of models such as the Dirac-type models, which have only two bands with distinct energies (each of which generically has a degeneracy),  $z({\bf k})$ must be zero.
In these cases, 
it is easy to show that the integrand in
the remaining terms are independent of the energy spectrum and are proportional to
$Tr \mathcal{F} \wedge \mathcal{F}$. Thus a direct and explicit evaluation of
$x({\bf k}) + y({\bf k})$ reveals eq. (F1).

\subsection{Topological invariance of $f$ upon smooth deformations of $G$}

For general band structures with three or more distinct energy levels, the proof of eq. (\ref{F1})
is more nontrivial and it is essential to exploit the topological invariance of $f$ to deformations that
preserve the band structure near the Fermi surface. Under a smooth deformation of $G$ to $G + \delta G$, the variation of $f$ is given by
\begin{align}
\label{deltaC2}
\delta f = \frac{i\pi^2}{15} \epsilon^{\mu \nu \rho \sigma \tau} \int \frac{d^4k d\omega}{(2\pi)^5}
\partial_\mu Tr[ (G^{-1} \delta G)(\partial_\nu G^{-1} G) (\partial_\nu G^{-1} G) (\partial_\nu G^{-1} G) (\partial_\nu G^{-1} G)].
\end{align}
Note that a deformation of the band structure that changes the location of the Fermi surface is not
a smooth deformation, because it can move the poles of $G$ from the upper half plane to the lower half plane (or vice versa),
and these are not small changes. Therefore, a small deformation of the band structure corresponds to a
small deformation of $G$ provided that it is away from the Fermi surface. Thus we can only deform the band structure if
we keep a small window near the Fermi surface unchanged. Observe that since the integrand in (\ref{deltaC2}) is a total
derivative, it will vanish if $\delta G$ vanishes at the boundaries of the integral. Therefore, we can deform
an arbitrary band structure (Fig. \ref{deformedBands}a) into a flat band dispersion (Fig. \ref{deformedBands}b), where the band structure is unchanged
in a small window near the Fermi surface. We may take the limit that the size of the window goes to zero at the end of
the calculation.

\begin{figure}[tb]
\centerline{
\includegraphics[width=4.5in]{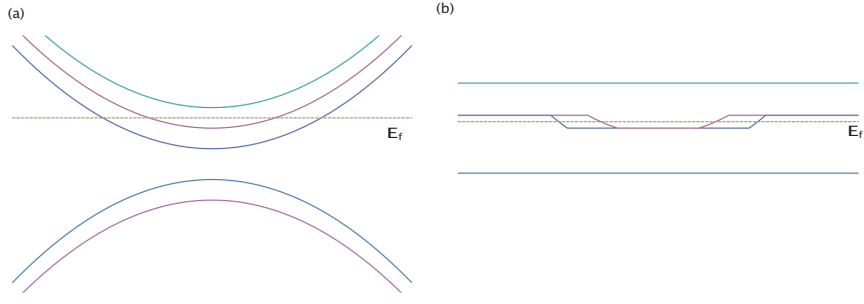}
}
\caption{
\label{deformedBands}
(a) Cross-section of a generic band structure, with the Fermi level lying within a
band. (b) Deformation of band structure in such a way that $f$ is left invariant.
The filled and empty bands are deformed to a flat
band, while the partially filled bands are deformed to flat bands everywhere
except a small window near the Fermi surface, where they are left unchanged.
}
\end{figure}

\subsection{Identities}

For the proof of eqn. (\ref{F1}) and (\ref{F2}), here we will prove the following identity:
\begin{align}
\label{CSconstraint}
\epsilon^{ijk} \sum_{P, \ga} \int_{FS} \mathcal{A}^{(M); P \ga}_i \mathcal{F}^{(M); \ga P}_{jk} = 0.
\end{align}
$FS$ stands for an arbitrary 3D Fermi surface in the 4D Brillouin zone. $\mathcal{F}^{(M)}$ is the curvature associated with $M$ bands, including $m$ bands below the Fermi surface labeled by $\ga$ and $M-m$ bands crossing the Fermi surface labeled by $P$.
Our proof will require us to be able to set $\xi_P({\bf k}) > \xi_\ga({\bf k})$ for all ${\bf k}$, so (\ref{CSconstraint}) is true only for
those bands for which this is possible.

First we observe that for an insulator,
\begin{align}
f =  \frac{1}{32 \pi^2} \int_{BZ} Tr \mathcal{F} \wedge \mathcal{F},
\end{align}
where $\mathcal{F}$ is the field strength of the Berry connection associated with the filled bands. This formula is
easy to prove in the case that the bands are all flat, with the empty/filled bands having energy $\xi_E$ or $\xi_F$,
respectively \cite{qi2008b}. For a general band structure, it is argued that $f$ is invariant under deformations of the band structure,
and so its value is given by deforming the energy dispersions to the flat band model. This means that the terms
in (\ref{C2simple}) that depend on energy must vanish, imposing various constraints on the properties of the Berry connections.
(\ref{CSconstraint}) is one of these constraints; here we will derive it, along with some others as well.

First we consider a band structure with flat energy dispersions $\xi_E$, $\xi_P$, and $\xi_F$, each of which may be multiply degenerate.
When $\xi_E$ and $\xi_P$ are empty, it is easy to show that
\begin{align}
x({\bf k}) + y({\bf k}) = &-6*2\pi i (A_{EFEF} + A_{PFPF} + A_{PFEF} + A_{EFPF})
\nonumber \\
& + 8 \pi i \sum_{n=2}^{\infty}  (-\xi_{EP}/\xi_{FE})^n (A_{PEPF} - A_{PEFE} - A_{PFEF} )
\end{align}
We have set
\begin{align}
A_{ABCD} = \epsilon_{ijkl} \sum_{(\alpha,\beta,\gamma,\delta) \in (A,B,C,D)} \mathcal{A}^{\alpha \beta}_i \mathcal{A}_j^{\beta \gamma} \mathcal{A}_k^{\gamma \delta} \mathcal{A}_l^{\delta \alpha}  ,
\end{align}
where the indices $A,B,C,D$ can refer to $P$, $E$, or $F$.
When the bands are all flat, $z({\bf k}) = 0$, so $f = -\frac{i\pi^2}{3} \frac{1}{(2\pi)^5} \int d^4k (x + y)$. Since the second terms (whose coefficients depend on
the energies) must vanish, it follows that
\begin{align}
\label{constraint1}
\int d^4k  (A_{PEPF} - A_{PEFE} - A_{PFEF}) = 0.
\end{align}

\begin{figure}[tb]
\centerline{
\includegraphics[width=4.5in]{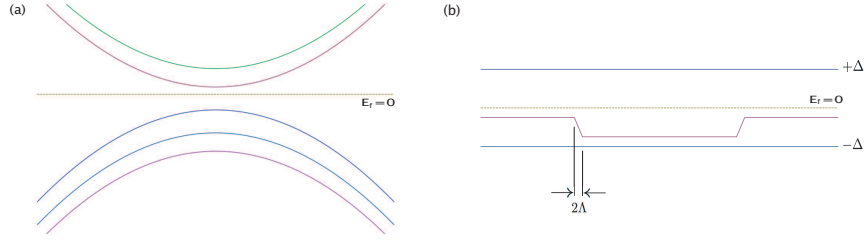}
}
\caption{
\label{deformedBands2}
(a) Cross-section of a generic band structure with $M$ filled bands. (b) Deformed band structure: the empty bands are
deformed to a flat band with energy $\Delta$, $m$ filled bands are deformed to flat bands with energy $-\Delta$, and $M-m$ filled bands
are deformed so that the energy is flat everywhere except in a small window of width $2\Lambda$ where it varies linearly, as discussed in the
main text.
}
\end{figure}

Now consider a general band structure for an arbitrary insulator that has $M$ filled bands.
We know that we can deform the bands however we like without changing $f$, as long as no band crosses the Fermi level during the deformation. Let us
deform them in the following way: consider a band structure where there are bands with
energy $\xi_E$ that are empty, $m$ bands with energy $\xi_F$ that are filled, and
$M - m$ bands with energy $\xi_P$ that are filled. $\xi_E$ and $\xi_F$ will be flat
bands, with energy $\pm \Delta$, respectively. $\xi_P$ will be flat everywhere
except a small region in the BZ, where it varies linearly (See Fig. \ref{deformedBands2}). More specifically,
pick a boundaryless 3-manifold, called $FS$, and suppose that $\xi_P$ varies linearly with distance from
$FS$ in some shell of thickness $2\Lambda$ around $FS$. In this case, $Z = \int_{BZ} d^4k z({\bf k}) $ is given by
\begin{align}
Z = &-i \epsilon^{ijkl} \sum_P \sum_{\alpha \gamma}^{'}
\int_{FS \pm \Lambda}  d^4k \partial_i \xi_P \xi_{P \alpha} \xi_{\gamma P} \xi_{\alpha \gamma} (\mathcal{A}_j^{\alpha P} \mathcal{A}_k^{P \ga} \mathcal{A}_l^{\ga \alpha} +
\mathcal{A}_j^{\alpha \ga} \mathcal{A}_k^{\ga P} \mathcal{A}_l^{P \alpha})
\int d\omega  d_\alpha^2 d_P^2 d_\gamma ,
\nonumber \\
 = &-i \epsilon^{ijkl} \sum_P \int_{FS \pm \Lambda}  d^4k d\omega \xi_{FP} \xi_{EF} \xi_{PE} \partial_i \xi_P d_P^2 d_E d_F (d_E - d_F)
\sum_{\beta \in E, \gamma \in F} ( \mathcal{A}_j^{\alpha P} \mathcal{A}_k^{P \ga} \mathcal{A}_l^{\ga \alpha}
+ \mathcal{A}_j^{\alpha \gamma} \mathcal{A}_k^{\gamma P} \mathcal{A}_l^{P \alpha}   )
\nonumber \\
 =& \epsilon^{ijkl} \sum_P \int_{FS \pm \Lambda}  d^4k d\omega \xi_{FP} \xi_{EF} \xi_{PE} \partial_i \xi_P d_P^2 d_E d_F (d_E - d_F)
\frac{1}{2} \sum_{\gamma \in F} ( \mathcal{A}_j^{P\gamma} \mathcal{F}_{kl}^{\gamma P} + \mathcal{A}_j^{\gamma P} \mathcal{F}_{kl}^{P \gamma}  )
\nonumber \\
=& \int_{FS \pm \Lambda} d^4k d\omega V^i \partial_i \xi_P \xi_{FP}
\xi_{EF} \xi_{PE} d_P^2 d_E d_F (d_E - d_F)
\nonumber \\
=&  2\pi i \int_{FS \pm \Lambda} d^4k V^i \partial_i \xi_P \xi_{PF}
\xi_{FE} \xi_{EP} [ \frac{1}{\xi_{FP}^2 \xi_{FE}^2} +
  \frac{1}{\xi_{EP}^2 \xi_{EF}^2} - \frac{1}{\xi_{PE}^2\xi_{PF}^2}  - \frac{2}{\xi_{PE}^3 \xi_{PF}}  ] ,
\end{align}
where we have set
$V^i({\bf k}) = \frac{1}{2} \sum_P \sum_{\gamma \in F} ( \mathcal{A}_j^{P\gamma} \mathcal{F}_{kl}^{\gamma P} + \mathcal{A}_j^{\gamma P} \mathcal{F}_{kl}^{P \gamma}  )$.
For the sake of being explicit, we set the energy dispersions to be:
\begin{align}
\xi_E &= - \xi_F = \Delta,
\nonumber
\end{align}
\begin{equation}
\xi_P({\bf k}) =  \left\{
  \begin{array}{lll}
    \frac{\Delta}{4 \Lambda} \kappa - \Delta/2 &\mbox{ if } & {\bf k} \in (FS \pm \Lambda)\\
    -\Delta/4 & \mbox{ if } & {\bf k} \in (BZ \backslash (FFS + \Lambda)) \\
    -3\Delta/4 & \mbox{ if } & {\bf k} \in (FFS - \Lambda)  \\
    \end{array} \right.
\end{equation}
Here we have set $\kappa$ to be the distance in the direction $\hat n$ normal to $FS$.
\begin{align}
\xi_{PF} &= \Delta (1 + \xi_P/\Delta ) = \Delta (1/2 + \kappa / 4\Lambda)
\nonumber \\
\xi_{EP} &= \Delta(1 - \xi_P/\Delta) = \Delta(3/2 - \kappa / 4 \Lambda)
\end{align}
\begin{align}
Z = 2\pi i \frac{\Delta}{4 \Lambda}\int_{FS \pm \Lambda} d^4k V^{\hat n} \xi_{PF} \xi_{FE} \xi_{EP} [ \frac{1}{\xi_{FP}^2 \xi_{FE}^2} +
  \frac{1}{\xi_{EP}^2 \xi_{EF}^2} - \frac{1}{\xi_{PE}^2\xi_{PF}^2}  - \frac{2}{\xi_{PE}^3 \xi_{PF}}  ].
\end{align}
After some simplification, this becomes
\begin{align}
Z = &-\pi i \int_{-1}^1 d \kappa' \int_{FS} d^3k_{||} V^{\hat n}(\Lambda \kappa', {\bf k}_{||}) (\frac{1}{2} + \frac{\kappa'}{4})(\frac{3}{2} - \frac{\kappa'}{4}) \times
\nonumber \\
& [\frac{1}{4 (1/2 + \kappa'/4)^2 } +  \frac{1}{4 (3/2 - \kappa'/4)^2 } - \frac{1}{ (1/2 + \kappa'/4)^2  (3/2 - \kappa'/4)^2} + \frac{2}{(3/2 - \kappa'/4)^3(1/2 + \kappa'/4)}],
\end{align}
where we have made a change of variables $\kappa' = \kappa /\Lambda$.
Thus, it is clear that
\begin{align}
\lim_{\substack{\Delta, \Lambda \rightarrow 0 \\ \Delta/\Lambda \text{ fixed}}} Z \propto \int_{FS} V \cdot \hat{n}
\end{align}
Now observe that
\begin{align}
\label{C2eqn}
f = \frac{1}{32\pi^2} \int d^4k Tr F \wedge F  + \delta f^{out} + \delta f^{in},
\end{align}
where $\delta f^{out}$ arises from the contribution of integrals outside of $FS \pm \Lambda$, while $\delta f^{in}$ arises from
contributions of integrals over $FS \pm \Lambda$:
\begin{align}
\delta f^{out} &=  \frac{-i \pi^2}{3} \frac{1}{(2\pi)^5} 8 \pi i \sum_{n=2}^{\infty}  (-\xi_{EP}/\xi_{FE})^n  \left(\int_{BZ \backslash (FFS + \Lambda)} +
\int_{FFS - \Lambda} \right) d^4k [ (A_{PEPF} - A_{PEFE} - A_{PFEF} ) ]
\nonumber \\
\delta f^{in} &= \frac{-i \pi^2}{3} \frac{1}{(2\pi)^5}  8 \pi i \sum_{n=2}^{\infty} (-1/\xi_{FE})^n \int_{FS\pm \Lambda} d^4k \xi_{EP}^n({\bf k}) [A_{PEPF} - A_{PEFE} - A_{PFEF} ] + Z
\end{align}
For any choice of $\Delta$, it must be the case that
\begin{align}
\delta f^{out} + \delta f^{in} = 0,
\end{align}
because $f$ is actually given by the first term of (\ref{C2eqn}).
First observe that
\begin{align}
\lim_{\substack{\Delta, \Lambda \rightarrow 0 \\ \Delta/\Lambda \text{ fixed}}} \delta f^{in} = \lim_{\substack{\Delta, \Lambda \rightarrow 0 \\ \Delta/\Lambda \text{ fixed}}} Z.
\end{align}
This is because in that limit, it is clear that
$\sum_{n=2}^{\infty} \int_{FS\pm \Lambda} d^4k[  (\xi_{EP}({\bf k})/\xi_{EF})^n (A_{PEPF} - A_{PEFE} - A_{PFEF} ) ]  \rightarrow 0$, because the
integrand is finite while the integration region becomes vanishingly small.
Furthermore,
\begin{align}
\lim_{\substack{\Delta, \Lambda \rightarrow 0 \\ \Delta/\Lambda \text{ fixed}}} \delta f^{out} \propto \sum_{n=2}^{\infty}  (-\xi_{EP}/\xi_{FE})^n  \int d^4k [ (A_{PEPF} - A_{PEFE} - A_{PFEF} ) ] = 0.
\end{align}
The last equality above follows from eqn. (\ref{constraint1}).
From the above equations, it follows that
\begin{align}
\lim_{\substack{\Delta, \Lambda \rightarrow 0 \\ \Delta/\Lambda \text{ fixed}}} Z \propto \int_{FS} V \cdot \hat{n} = 0
\end{align}
Now observe that
\begin{align}
\int_{FS} V \cdot \hat{n} \propto \epsilon^{ijk} \sum_{P, \ga} \int_{FS} \mathcal{A}^{(M); P \ga}_i \mathcal{F}^{(M); \ga P}_{jk} = 0,
\end{align}
which proves the identity (\ref{CSconstraint}).

\subsection{Proof of Eqn. (\ref{F1})}

Armed with the identity (\ref{CSconstraint}), here we will prove eqn. (\ref{F1}) by deforming an
arbitrary band structure -- in the case where the chemical potential lies inside a band -- in such a way
that keeps $f$ unchanged while making it amenable to calculation.

Consider the following deformation of the band structure, which keeps $f$ unchanged because it
does not deform the partially filled bands in a window near the Fermi surface (see Fig. \ref{deformedBands3}).
As in the previous
section, we will use the variable $\ka$ to denote the distance from the Fermi surface, and $k_{||}$
to be the directions parallel to the Fermi surface. Choose some small energy $\Delta$ near zero
so that there exists $\Lambda_1$ and $\Lambda_2$ where
$\xi_P(\ka = \Lambda_1) = \Delta$ and $\xi_P(\ka = \Lambda_2) = -\Delta$. Deform
the empty and filled bands to have a flat dispersion, with energy $\pm \Delta$.
Then, deform the partially filled band for $\ka \geq \Lambda_1$ to have a flat dispersion
with energy $\Delta$. Finally deform the partially filled band for $\ka \leq \Lambda_2$ to
have a flat dispersion with energy $-\Delta$. For $\Lambda_2 \leq \ka \leq \Lambda_1$,
the dispersion of the partially filled band is unchanged. The region where
$\Lambda_2 \leq \ka \leq \Lambda_1$ will be denoted as $FS \pm \Lambda$.

Now we can split up the integrals over momentum space into a sum of integrals:
\begin{align}
f = -\frac{i\pi^2}{3} \frac{1}{(2\pi)^5} [ ( X_{out} + Y_{out} + Z_{out}) + (X_{in} + Y_{in} + Z_{in})],
\end{align}
where $X_{out} = \int_{BZ \backslash (FS \pm \Lambda)} d^4k x({\bf k})$, $X_{in} = \int_{FS \pm \Lambda} d^4k x({\bf k})$,
and analogously for $Y$ and $Z$. $Z_{out} = 0$ because it depends on $\partial \xi_\alpha$, and
\begin{align}
-\frac{\pi^2}{3} \frac{1}{(2\pi)^5} (X_{out} + Y_{out} + Z_{out}) =
 \frac{1}{32 \pi^2} \int_{\ka > \Lambda_1} d^4k \epsilon^{ijkl} Tr[\mathcal{F}^{(M)}_{ij}\mathcal{F}^{(M)}_{kl}] +
\frac{1}{32 \pi^2} \int_{\ka < \Lambda_2} d^4k \epsilon^{ijkl} Tr[\mathcal{F}^{(M+P)}_{ij}\mathcal{F}^{(M+P)}_{kl}] ,
\end{align}
where $\mathcal{F}^{(M)}$ is the Berry curvature for the $M$ filled bands outside of $FS$, while
 $\mathcal{F}^{(M + P)}$ is the Berry curvature for the $M$ filled bands and the $P$ partially filled bands whose
Fermi surface $FS$ we are considering.

\begin{figure}[tb]
\centerline{
\includegraphics[width=4.5in]{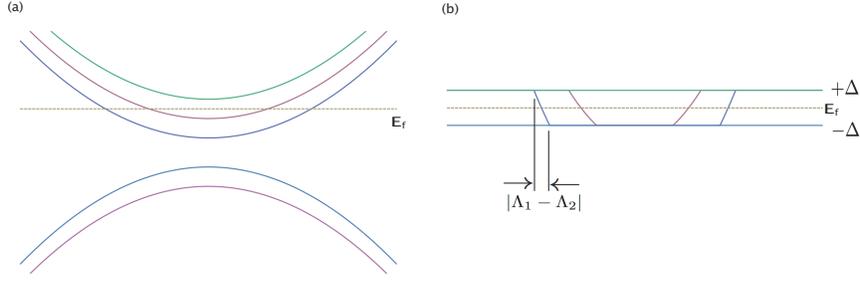}
}
\caption{
\label{deformedBands3}
(a) Cross section of generic band structure with partially filled bands.
(b) Deformed band structure. The bands are deformed to flat bands with energy $\pm \Delta$ relative to
the Fermi level everywhere outside of a small window near the Fermi surface. Inside the small window near
the Fermi surface, the bands are unchanged. The limit $\Delta \rightarrow 0$ is taken at the end of the
calculation.
}
\end{figure}

Observe that in the limit $\Delta \rightarrow 0$, $X_{in} + Y_{in} \rightarrow 0$ because the size of the region of
integration goes to zero while the integrand stays finite. In the following, we will show that
\begin{align}
\label{Zlim}
\lim_{\Delta \rightarrow 0} Z \propto  \epsilon^{ijk} \sum_{P, \ga} \int_{FS} \mathcal{A}^{(M); P \ga}_i \mathcal{F}^{(M); \ga P}_{jk} = 0,
\end{align}
which will show that in the case of a single Fermi surface,
\begin{align}
f = \frac{1}{32 \pi^2} \int_{BZ \backslash FFS} d^4k \epsilon^{ijkl} Tr[\mathcal{F}^{(M)}_{ij}\mathcal{F}^{(M)}_{kl}] +
\frac{1}{32 \pi^2} \int_{FFS} d^4k \epsilon^{ijkl} Tr[\mathcal{F}^{(M+P)}_{ij}\mathcal{F}^{(M+P)}_{kl}] .
\end{align}
The generalization to multiple non-intersecting Fermi surfaces is straightforward.

Now consider $Z$:
\begin{align}
Z = &-i \epsilon^{ijkl} \sum_P \sum_{\alpha \gamma}^{'}
\int_{FS \pm \Lambda}  d^4k \partial_i \xi_P \xi_{P \alpha} \xi_{\gamma P} \xi_{\alpha \gamma} (\mathcal{A}_j^{\alpha P} \mathcal{A}_k^{P \ga} \mathcal{A}_l^{\ga \alpha} +
\mathcal{A}_j^{\alpha \ga} \mathcal{A}_k^{\ga P} \mathcal{A}_l^{P \alpha})
\int d\omega  d_\alpha^2 d_P^2 d_\gamma ,
\nonumber \\
 = &-i \epsilon^{ijkl} \sum_P \int_{FS \pm \Lambda}  d^4k d\omega \xi_{FP} \xi_{EF} \xi_{PE} \partial_i \xi_P d_P^2 d_E d_F (d_E - d_F)
\sum_{\beta \in E, \gamma \in F} ( \mathcal{A}_j^{\alpha P} \mathcal{A}_k^{P \ga} \mathcal{A}_l^{\ga \alpha}
+ \mathcal{A}_j^{\alpha \gamma} \mathcal{A}_k^{\gamma P} \mathcal{A}_l^{P \alpha}   )
\nonumber \\
 =& \epsilon^{ijkl} \sum_P \int_{FS \pm \Lambda}  d^4k d\omega \xi_{FP} \xi_{EF} \xi_{PE} \partial_i \xi_P d_P^2 d_E d_F (d_E - d_F)
\frac{1}{2} \sum_{\gamma \in F} ( \mathcal{A}_j^{P\gamma} \mathcal{F}_{kl}^{\gamma P} + \mathcal{A}_j^{\gamma P} \mathcal{F}_{kl}^{P \gamma}  )
\nonumber \\
=& \int_{FS \pm \Lambda} d^4k d\omega V^i \partial_i \xi_P \xi_{FP}
\xi_{EF} \xi_{PE} d_P^2 d_E d_F (d_E - d_F)
\nonumber \\
=&  2\pi i\int_{FS \pm \Lambda} d^4k V^i \partial_i \xi_P \xi_{PF}
\xi_{FE} \xi_{EP} [ \frac{1}{\xi_{FP}^2 \xi_{FE}^2} +
  \frac{1}{\xi_{EP}^2 \xi_{EF}^2} - \frac{1}{\xi_{PE}^2\xi_{PF}^2} - \frac{2}{\xi_{PE} \xi_{PF}^3} +
\nonumber \\
& 2n_P ( \frac{1}{\xi_{PE} \xi_{PF}^3} - \frac{1}{\xi_{PE}^3 \xi_{PF}} ) ]  ,
\end{align}
We have:
\begin{align}
\xi_{PF} &= \Delta (1 + X)
\nonumber \\
\xi_{EF} &= 2\Delta
\nonumber \\
\xi_{EP} &= \Delta (1 - X),
\nonumber \\
X &=\xi_P/\Delta.
\end{align}
Thus
\begin{align}
Z \propto \frac{1}{\Delta} \int_{FS \pm \Lambda} d^4k V \cdot \partial \xi_P I({\bf k}, \Lambda_1),
\end{align}
where
\begin{align}
I({\bf k}, \Lambda_1) = (1 + X) (1 - X)[ &\frac{1}{4 (1 + X)^2} +
\frac{1}{4(1-X)^2} - \frac{1}{(1-X)^2 (1 + X)^2} + \frac{2}{(1-X)(1+X)^3}
+2n_P ( \frac{1}{(1-X)^3(1 + X)}
\nonumber \\
& - \frac{1}{(1-X)(1 + X)^3} ) ].
\end{align}
Simplifying:
\begin{align}
I({\bf k}, \Lambda_1) = \left\{
  \begin{array}{lll}
      \frac{(1 - X)(X+3)}{2(1+X)^2}  &\mbox{ if } & X({\bf k}) > 0\\
     \frac{(1 +X)(3- X)}{2(1-X)^2} & \mbox{ if } & X({\bf k}) \leq 0 \\
    \end{array} \right.
\end{align}
In the limit $\Delta \rightarrow 0$, which by definition also takes $\Lambda_i \rightarrow 0$ while
keeping $\Delta/\Lambda_i$ fixed, we see that the size of the integration region is proprotional to
$\Lambda_1 - \Lambda_2$, while $I({\bf k}, \Lambda_1)$ is non-singular. It follows that
\begin{align}
\lim_{\Delta \rightarrow 0} Z \propto \int_{FS} V \cdot \partial \xi_P \propto \int_{FS} V \cdot \hat{n},
\end{align}
where the latter proportionality follows because $\xi_P$ has non-zero slope at the
Fermi surface only in the direction normal to the Fermi surface. From the definition of
$V$, (\ref{Zlim}) follows.

\section{Proof of Eqn. (\ref{F2}) }

In the last section, we proved the highly non-trivial relation
\begin{align}
f = \frac{1}{32\pi^2}\int d^4{\bf k}{\rm  Tr}\left( \mathcal{F}_{ab} \mathcal{F}_{cd}\right) \epsilon^{abcd}.
\end{align}
The dependence on the chemical potential is implicit in the definition of the field strength $\mathcal{F}({\bf k})$. $\mathcal{F}({\bf k})$ is defined
to be the field strength of the Berry connection for the bands that are filled at the point ${\bf k}$ in the four-dimensional
Brillouin Zone. Depending on the location of the Fermi surface(s), the number of filled bands will be different at different
${\bf k}$-points. This is not the typical formula for a second Chern invariant.

Here we prove a more mathematical result:
\begin{align}
\label{C2fracSupp}
f \text{ mod } 1 =  \frac{1}{8 \pi^2}\sum_i \int_{FS_i} \mathcal{L}_{CS}(\mathcal{A}^{P_i}),
\end{align}
where $\mathcal{A}^{(P_i)}$ is the Berry connection involving only the partially filled bands that are
part of the $i$th Fermi surface $FS_i$.

In order to prove this, let us first concentrate on the case of a single Fermi surface. We may rewrite $f$ as
\begin{align}
f = \frac{1}{32\pi^2}\int_{BZ \backslash FFS} {\rm  Tr}\left( \mathcal{F}^{(0)}_{ab} \mathcal{F}^{(0)}_{cd}\right) \epsilon^{abcd} +
\frac{1}{32\pi^2}\int_{FFS} {\rm  Tr}\left( \mathcal{F}^{(1)}_{ab} \mathcal{F}^{(1)}_{cd}\right) \epsilon^{abcd},
\end{align}
where $FFS$ indicates the filled Fermi sea, \it ie \rm the region in momentum space inside the Fermi surface, while
$BZ \backslash FFS$ indicates the region outside of the filled Fermi sea. $\mathcal{F}^{(0)}$ involves the completely filled bands,
while $\mathcal{F}^{(1)}$ involves the completely filled bands and the partially filled bands.

Now observe that $F_{ab} F_{cd} \epsilon^{abcd}$ is locally a total derivative:
\begin{align}
F_{ab} F_{cd} \epsilon^{abcd} = 4 \partial_a (A_b \partial_c A_d  + \frac{2i}{3} A_b A_c A_d) \epsilon^{abcd},
\end{align}
which implies:
\begin{align}
f \text{ mod } 1 =  \frac{1}{8 \pi^2} \int_{FS} [\mathcal{L}_{CS}(\mathcal{A}^{(1)}) - \mathcal{L}^{CS}(\mathcal{A}^{(0)}) ],
\end{align}
where $\mathcal{A}^{(0)}$ is the non-Abelian Berry connection for the completely filled bands, $\mathcal{A}^{(1)}$ is the
non-Abelian Berry connection for the completely and partially filled bands, and
$\mathcal{L}^{CS}(A) = \epsilon^{\mu \nu \lambda} \rm{ Tr } [A_\mu \partial_\nu A_\lambda + \frac{2i}{3} A_\mu A_\nu A_\lambda ]$
is the Chern-Simons form. Since the Chern-Simons form is non-linear, this subtraction has a cross-term:
\begin{align}
f \text{ mod } 1 =  \frac{1}{8 \pi^2} \int_{FS} [\mathcal{L}_{CS}(\mathcal{A}^{(P)}) + \epsilon^{ijk} \sum_{\gamma \in F,P} \mathcal{A}^{P\gamma}_i \mathcal{F}_{jk}^{(1);\gamma P} ].
\end{align}
The second term vanishes by Eq. (\ref{CSconstraint}), so that we obtain
\begin{align}
f \text{ mod } 1 =  \frac{1}{8 \pi^2} \int_{FS} \mathcal{L}_{CS}(\mathcal{A}^{(P)}).
\end{align}

The generalization to multiple Fermi surfaces is straightforward and yields (\ref{C2fracSupp}).

\section{$f$ for Dirac -type Models}

Consider the Hamiltonian
\begin{align}
H({\bf k}) = \epsilon({\bf k}) \mathbb{I} + d_a \Gamma^a,
\end{align}
where $\Gamma^a$ for $a = 1,...,5$ are the $4\times 4$ $\Gamma$
matrices; their relation to the physical properties of Bi$_2$Se$_3$
are described in \cite{LQ1022}.
The eigenvalues of this Hamiltonian are $\xi_{\pm}({\bf k}) = \epsilon({\bf k}) \pm |d({\bf k})|$; each band is doubly degenerate.
$P_{\pm} = \frac{1}{2}(1 \pm \hat d \cdot \Gamma)$ projects onto the bands with energies $\xi_{\pm}$. The Green's function is:
\begin{align}
G({\bf k}, \omega) &= \frac{P_+}{\omega - \xi_+ +i \delta \sgn(\xi_+)} + \frac{P_-}{\omega - \xi_- + i\delta \sgn(\xi_-)}
\nonumber \\
&= \frac{\omega - (\xi_+ + \xi_-)/2 + i \delta [\sgn(\xi_+) + \sgn(\xi_-)]/2 + \hat d \cdot \Gamma [(\xi_+ - \xi_-)/2 + i \delta (\sgn(\xi_-) - \sgn(\xi_+))/2]}{(\omega - \xi_+ + i \delta \sgn(\xi_+))[\omega - \xi_- + i\delta \sgn(\xi_-)]}
\nonumber \\
&= \frac{\tilde{\omega} + \tilde{d} \cdot \Gamma}{(\omega - \xi_+ + i\delta \sgn(\xi_+))(\omega - \xi_- + i\delta \sgn(\xi_-))},
\end{align}
where $\tilde{\omega} = \omega - (\xi_+ + \xi_-)/2 + i \delta [\sgn(\xi_+) + \sgn(\xi_-)]/2$ and
$\tilde{d} = \hat{d} [(\xi_+ - \xi_-)/2 + i \delta (\sgn(\xi_-) - \sgn(\xi_+))/2]$.
Substituting into the formula gives
\begin{align}
f =& -\frac{\pi^2}{3} \epsilon_{ijkl} \int \frac{d^4k d\omega}{(2\pi)^5} \frac{1}{[(\omega - \xi_+({\bf k}) + i \delta \sgn(\xi_+))(\omega - \xi_-({\bf k}) +
i \delta \sgn(\xi_-))]^5}
\nonumber \\
& \times Tr[ (\tilde{\omega} + \tilde{d} \cdot \Gamma)^2 \partial_i (\epsilon({\bf k}) \mathbb{I} +
d_a \Gamma^a ) (\tilde{\omega} + \tilde{d} \cdot \Gamma) \partial_j(\epsilon({\bf k}) \mathbb{I} + d_a \Gamma^a)
(\tilde{\omega} + \tilde{d} \cdot \Gamma) \partial_k (\epsilon({\bf k}) \mathbb{I} + d_a \Gamma^a)
(\tilde{\omega} + \tilde{d} \cdot \Gamma) \partial_l (\epsilon({\bf k}) \mathbb{I} + d_a \Gamma^a) ]
\end{align}
Observe that any term with two or more factors of $\partial \epsilon$ will clearly vanish under the epsilon tensor.
The term with one factor of $\partial \epsilon$ can be shown to vanish.
The term with no factors of $\partial \epsilon$ gives:
\begin{align}
f = &-\frac{\pi^2}{3} \epsilon_{ijkl} \int \frac{d^4k d\omega}{(2\pi)^5}
\frac{\partial_i d_a \partial_j d_b \partial_k d_c \partial_l d_d}{[(\omega - \xi_+ + i\delta \sgn(\xi_+))(\omega - \xi_- + i\delta \sgn(\xi_-))]^5}
\nonumber \\
&\times Tr[ (\tilde{\omega} + \tilde{d} \cdot \Gamma)^2 \Gamma^a  (\tilde{\omega} + \tilde{d} \cdot \Gamma) \Gamma^b)
(\tilde{\omega} + \tilde{d} \cdot \Gamma) \partial_k \Gamma^c (\tilde{\omega} + \tilde{d} \cdot \Gamma) \Gamma^d ]
\end{align}
The trace simplifies to $- 4\epsilon_{tabcd} d^t (\tilde{\omega}^2 - \tilde{d}^2)^2$. Thus:
\begin{align}
f = &4\frac{\pi^2}{3} \epsilon_{ijkl} \int \frac{d^4k d\omega}{(2\pi)^5}
\epsilon_{tabcd} d^t\partial_i d_a \partial_j d_b \partial_k d_c \partial_l d_d
\frac{1}{[(\omega - \xi_+ + i\delta \sgn(\xi_+))(\omega - \xi_- + i\delta \sgn(\xi_-))]^3}
\end{align}
If the chemical potential lies in the conduction band, this is:
\begin{align}
f &= -\frac{\pi^2 i}{4} \epsilon_{ijkl} \int_{BZ \backslash FFS} \frac{d^4k}{(2\pi)^4}
\epsilon_{tabcd} \frac{d^t\partial_i d_a \partial_j d_b \partial_k d_c \partial_l d_d}{|d|^5}
\nonumber \\
&= -\frac{3}{8 \pi^2}\int_{BZ \backslash FFS} d^4k
\epsilon_{tabcd} \hat{d}^t\partial_x \hat{d}_a \partial_y \hat{d}_b \partial_z \hat{d}_c \partial_\theta \hat{d}_d
\end{align}
If the chemical potential lies in the valence band, we have:
\begin{align}
f = -\frac{3}{8 \pi^2}\int_{FFS} d^4k
\epsilon_{tabcd} \hat{d}^t\partial_x \hat{d}_a \partial_y \hat{d}_b \partial_z \hat{d}_c \partial_w \hat{d}_d
\end{align}

For Bi$_2$Se$_3$ with a surface T-breaking field, we have
\begin{align}
\epsilon({\bf k}) &= C_0 + C_1 k_Z^2 + C_2 k_{||}^2,
\nonumber \\
d_1({\bf k}) &= \mathcal{A}_0 k_y
\nonumber \\
d_2({\bf k}) &= - \mathcal{A}_0 k_x
\nonumber \\
d_3({\bf k}) &= m_0 \sin \theta
\nonumber \\
d_4({\bf k}) &= \mathcal{B}_0 k_z
\nonumber \\
d_5({\bf k}) &= -M_0 \cos \theta + M_1k_z^2 + M_2 k_{||}^2,
\end{align}
where we have only kept terms up to quadratic order in $k$. The
parameters are given in Ref. \cite{LQ1022}; we list them
in Table \ref{param} for completeness.

\begin{table}
\begin{tabular}{lc}
\hline
$\mathcal{A}_0$ (eV \AA) & 3.33 \\
$\mathcal{B}_0$ (eV \AA) & 2.26 \\
$C_0$ (eV) & -0.0083 \\
$C_1$ (eV \AA$^2$) & 5.74 \\
$C_2$  (eV \AA$^2$) & 30.4 \\
$M_0$ (eV) & -0.28  \\
$M_1$  (eV \AA$^2$) & 6.86 \\
$M_2$  (eV \AA$^2$) & 44.5 \\
\hline
\end{tabular}
\caption{
\label{param}
Parameters of the model Hamiltonian for Bi$_2$Se$_3$.
}
\end{table}

\section{Effect of top gating at the surface of a doped topological insulator }

Here we would like to develop a calculation of the effect of gating at the surface of a doped topological insulator. 

We assume that the chemical potential is slightly above the bottom of the conduction band, and that the conduction band bends upward
so that at the surface, the chemical potential lies in the bulk band gap. We also assume that a gap $m$ has opened in the Dirac
surface states and that the chemical potential is always inside of this gap. 

In the simplest depletion length approximation, we assume that within a depletion length $z_d$ from the surface, there are no
carriers; if the bulk doping charge density is $-en_b$, where $n_b$ is
the bulk carrier density, this means that there is a constant positive charge density $en_b$ 
in the depletion region. In order to maintain overall charge neutrality, there must be a compensating negative charge near the surface. 
Let us assume that all of this compensating negative charge comes from the application of a top gate. In practice
this is not entirely true, but it may be a good approximation. With such an approximation, we have:
\begin{align}
V_g C = e n_b z_d,
\end{align}
where $C$ is the capacitance per unit area of the region between the metallic top gate and the surface of the material. 

This negative charge is accumulated in a metal gate a distance $d$ from the surface of the material. Thus, the profile of
the charge density is:
\begin{align}
\rho(z) = -C V_g\delta(z+d) + e n_b z, \;\;\;\;\; z \leq z_d
\end{align}
Note that $z= 0$ is the surface of the material, and $z = - d$ is the position of the top gate. The potential $V(z)$ is therefore given by
\begin{align}
d^2 V/dz^2 = - \rho(z)/(\epsilon(z) \epsilon_0),
\end{align}
where $\epsilon(z) = \epsilon_{g}$ for $-d < z < 0$ is the dielectric constant of the dielectric that is in between the metal gate and
the surface of the material, and $\epsilon(z) = \epsilon$ for $z > 0$ is the dielectric constant of the topological insulator. 

We are interested in $V(z)$ for $z > 0$. For simplicity, let us assume that $d \approx 0$. The electric field in the region 
$0 \leq z \leq z_d$ is:
\begin{align}
E(z) = - \frac{CV_g}{2\epsilon \epsilon_0}  + \frac{en_b}{2\epsilon \epsilon_0} z.
\end{align}
The potential is then
\begin{align}
V(z) =  -\int_{z_d}^z E(z) = -\frac{C V_g}{4 \epsilon \epsilon_0} z_d (1-\frac{z}{z_d})^2 = -\frac{e n_g^2}{4 \epsilon \epsilon_0 n_b} (1-\frac{z n_b}{n_g})^2 
\end{align}
The potential energy associated with this is $eV(z)$.  Note that $z_d = V_g C/en_b = n_g/n_b$, where $n_g$ is the number
density of carriers on the top gate. 

Thus, we would like to fix the penetration depth of the magnetization, $\xi$, the gap of the Dirac surface states $m$ (which is induced
by the magnetization), and the bulk density $n_b$, and then study the surface transverse conductance as a function
of the gate voltage $V_g$. For Bi$_2$Se$_3$, we take $\epsilon \approx 100$, so $\epsilon \epsilon_0 \approx 8.85 10^{-12} F/cm$.

\end{widetext}

\end{document}